\documentclass{ws-procs961x669}            
\begin{document}
\title{A glimpse into Feynman's contributions to the debate on the foundations of quantum mechanics}

\author{M. Di Mauro}

\address{Dipartimento di Matematica, Universit\`a di Salerno,
Via Giovanni Paolo II,\\
Fisciano (SA), 84084, Italy\\
E-mail: madimauro@unisa.it}

\author{S. Esposito$^*$ and A. Naddeo$^{**}$}

\address{INFN, Sezione di Napoli, C. U. Monte S. Angelo, Via Cinthia,\\
Napoli, 80125, Italy\\
$^*$E-mail: sesposit@na.infn.it\\
$^{**}$E-mail: anaddeo@na.infn.it}

\begin{abstract}
The broad debate on foundational issues in quantum mechanics,
which took place at the famous 1957 Chapel Hill conference on
\textit{The Role of Gravitation in Physics}, is here critically
analyzed with an emphasis on Richard Feynman's contributions. One
of the most debated questions at Chapel Hill was whether the
gravitational field had to be quantized and its possible role in
wave function collapse. Feynman's arguments in favor of the
quantization of the gravitational field, based essentially on a
series of gedanken experiments, are here discussed. Then the
related problem of the wave function collapse, for which Feynman
hints to decoherence as a possible solution, is discussed.
Finally, another topic is analyzed, concerning the role of the
observer in a closed Universe. In this respect, Feynman's
many-worlds characterization of Everett's approach at Chapel Hill
is discussed, together with later contributions of his, including
a kind of Schr\"{o}dinger's cat paradox, which are scattered
throughout the 1962-63 Lectures on Gravitation. Philosophical
implications of Feynman's ideas in relation to foundational issues
are also discussed.
\end{abstract}


\bodymatter

\section{Introduction: the Chapel Hill conference}\label{sec1}
Richard Feynman's most famous contribution in quantum theory is
undoubtedly his celebrated path integral approach
\cite{Feynman1948}. His contributions to the debate on
interpretational issues are instead much less well known
\cite{Zeh,noi}. The first such contributions took place in the
wide discussions which characterized the 1957 Chapel Hill
conference \cite{ChapelHill}, whose title was: \emph{The Role of
Gravitation in Physics}. The Chapel Hill conference, which played
a key role in triggering the so called \emph{Renaissance} of
general relativity \cite{BlumLalliRenn} was of capital importance
in establishing the future research lines in the field. Broadly,
the main tracks were\cite{BergmannRMP}: classical gravity, quantum
gravity, and the classical and quantum theory of measurement (as a
link between the previous two topics). In particular, foundational
quantum issues were widely discussed by researchers attending the
conference. The main motivation for addressing them came from the
fundamental question of whether the gravitational field had to be
quantized, like other fundamental fields, or not. As its title
declares, the ambitious goal of the conference was the merging of
general relativity with the rest of physics and, especially, with
the world of elementary particle physics, which was ruled by
quantum mechanics, hence it was (and it still is, of course) very
important to understand physics in regimes where both gravity and
quantum mechanics become important. This was clearly stated by
Peter G. Bergmann in the opening session of the second half of the
conference, which was in fact focused on problems within the
quantum domain (Ref. \citenum{ChapelHill}, p. 165):
\begin{quote}
Physical nature is an organic whole, and various parts of
physical theory must not be expected to endure in ``peaceful
coexistence.'' An attempt should be made to force separate
branches of theory together to see if they can be made to merge,
and if they cannot be united, to try to understand why they clash.
Furthermore, a study should be made of the extent to which
arguments based on the uncertainty principle force one to the
conclusion that the gravitational field must be subject to quantum
laws: (a) Can quantized elementary particles serve as sources for
a classical field? (b) If the metric is unquantized, would this
not in principle allow a precise determination of both the
positions and velocities of the Schwarzschild singularities of
these particles?
\end{quote}
The answer to these questions was expected to lead to a novel
perspective on the ordinary notions of space and time, by
introducing uncertainty relations for quantities such as distances
and volumes. Physicists hoped that, as a result, the divergences
which plagued quantum field theory would get suppressed by the
gravitational field. Along with the formal discussion on the main
approaches to the quantization of the gravitational field (namely
the canonical, the functional integral and the covariant
perturbative approaches), which would have been developed in the
following decades, a lot of time was devoted to conceptual issues,
especially in discussion sessions. A broad debate arose on the
problem of quantum measurement, the main conceptual question
being:
\begin{quote}
What are the limitations imposed by the quantum theory on the
measurements of space-time distances and curvature? (Ref.
\citenum{ChapelHill}, p. 167)
\end{quote}
or, equivalently
\begin{quote}
What are the quantum limitations imposed on the measurement of the
gravitational mass of a material body, and, in particular, can the
principle of equivalence be extended to elementary particles?
(Ref. \citenum{ChapelHill}, p. 167)
\end{quote}
As the editors of the written records of the
conference\cite{ChapelHill} emphasized, an answer to the above
questions could not be simply found in dimensional arguments,
since the Planck mass does not set a lower limit to the mass of a
particle whose gravitational field can be measured. Indeed, a
simple argument shows that the gravitational field of \emph{any}
mass can in principle be measured, thanks to the ``long tail'' of
the Newtonian force law (Ref. \citenum{ChapelHill}, pp. 167-8).

The great part of the discussions of interest to us here developed
in Section VIII of the conference (Ref. \citenum{ChapelHill}, pp.
243-60), where the focus was mainly on the contradictions
eventually arising in the logical structure of quantum theory if
gravity quantization is not assumed, but in fact many even more
foundamental issues for quantum physics in general were touched as
well. Feynman greatly contributed to these discussions, proposing
several gedanken experiments in order to argue in favor of the
necessity of gravitational quantization, and hinting to
decoherence as a viable solution to the problem of wave function
collapse. His contributions in fact triggered a wide debate on the
quantum measurement problem and on the existence and meaning of
macroscopic quantum superpositions. Further, in the subsequent
session of the conference, which was the closing one (Ref.
\citenum{ChapelHill}, pp. 263-278), he gave a ``many-worlds''
characterization of Hugh Everett III's relative state
interpretation of quantum mechanics \cite{Everett}, which had just
been described for the first time by Everett's advisor, John A.
Wheeler. It is interesting to notice that, in fact, the Chapel
Hill conference was one of the few places where Feynman was
directly involved in discussions about the foundations and
interpretation of quantum mechanics \cite{Zeh}.

The Chapel Hill discussions played a pivotal role in triggering
subsequent research on foundational issues, mainly on the quantum
measurement problem and on the various possible interpretations of
quantum mechanics, with an emphasis on the consequences of
Everett's one. In this respect it is worth mentioning further
contributions by Feynman, who despite not being often involved
with foundational quantum issues, nonetheless kept thinking deeply
on them also in the following years, especially when working on
gravity, as hinted for example in a letter written in 1961 to
Viktor Weisskopf \cite{WeisskopfLetter}:
\begin{quote}
How can we experimentally verify that these waves are quantized?
Maybe they are not. Maybe gravity is a way that quantum mechanics
fails at large distances.
\end{quote}
as well as in several suggestions and considerations scattered
throughout his 1962-63 graduate lectures on gravitation
\cite{Feynman:1996kb}.

In this paper, we historically and critically analyze Feynman's
contributions to the debate about the foundations of quantum
mechanics and gravity, with an emphasis on the Chapel Hill
discussions, which constitute the main source about his thoughts
on the matter. In particular, Section 2 deals with the problem of
the quantization of the gravitational field while in Sections 3
and 4 the focus is on wave function collapse following a
measurement and on Everett's relative state interpretation of
quantum mechanics. Finally our concluding remarks are summarized
in Section 5.

\section{Quantization of the gravitational field: gedanken experiments}

In this Section we report on Feynman's arguments in favor of the
quantization of the gravitational field. As a matter of fact,
Feynman believed that nature cannot be half classical and half
quantum and shared Bergmann's general ideas, as recalled by
himself in some concluding remarks made at the end of the
conference:
\begin{quote}
The questions raised in the last three days have to do with the
relation of gravity to the rest of physics. We have gravity -
electrodynamics - quantum theory - nuclear physics - strange
particles. The problem of physics is to put them all together. The
original problem after the discovery of gravity was to put gravity
and electrodynamics together since that was essentially all that
was known. Therefore, we had the unified field theories. After
quantum theory one tries to quantize gravity (Ref.
\citenum{ChapelHill}, p. 272).
\end{quote}
In particular, for Feynman gravity, like the other fundamental
interaction that we experience at the macroscopic, classical
level, i.e. electromagnetism, has a quantum foundation (see e.g.
Refs. \citenum{DeLuca:2019ija,DiMauro:2020bpd,FeynmanHughes2}).
Coherently with this, he developed an approach to quantum gravity
(whose first hints were given at Chapel Hill, cf. Ref.
\citenum{ChapelHill}, pp. 271-276) which was characterized by
being fully quantum from the beginning
\cite{Feynman:1996kb,Feynman:1963ax}.

Feynman's arguments in favor of the quantization of gravity were
totally different from the usual dimensional arguments, which rely
on the assumption that gravity has to dominate over all other
interactions at the Planck scale. According to him quantum effects
involving gravity should be subject to probing without invoking
such high scales. He provided further evidence to support this
claim by putting forward thought experiments designed to show
that, if quantum mechanics is required to hold for objects massive
enough to produce a detectable gravitational field (the opposite
would require a modification of quantum mechanics), then the
gravitational field has to be quantized, in order to avoid
contradictions. This view supported the hope that quantization of
the metric would help taming the divergences present in quantum
field theory and, in this way, it was relevant also for the theory
of elementary particles. Let us thus retrace the discussion in
more detail.

Section VIII began with Thomas Gold stating that, in the absence
of phenomena not explainable by classical gravity, the only way to
argue in favor of quantization of gravity is the presence of
logical contradictions, of which he was not convinced (Ref.
\citenum{ChapelHill}, pp. 243-244). Bryce S. DeWitt pointed out
that difficulties may arise in the presence of quantized matter
fields depending on the choice of the quantum expectation value of
the stress-energy tensor as the source of the gravitational field
(Ref. \citenum{ChapelHill}, p. 244). Indeed a measurement may
change this expectation value, which implies a change of the
gravitational field itself. The classical theory of gravitation is
valid because fluctuations are negligible at the scale where
gravitational effects become sizeable. At this point Feynman put
forward his first thought experiment, which was a variant of the
two-slit diffraction experiment, in which a mass indicator has
been put behind the two-slit wall. Within a space-time region
whose linear dimensions are of order $L$ in space and ${L}/{c}$ in
time, the uncertainty on the gravitational potential (divided by
$c^2$ so that it is dimensionless and thus homogeneous to a
metric) is in general $\Delta
g=\sqrt{\frac{hG}{c^3L^2}}=\frac{L_P}{L}$, where
$L_P=\sqrt{\frac{hG}{c^3}}$ is the Planck length.\footnote{This
follows from an argument given by Wheeler in Ref.
\citenum{ChapelHill}, pp. 179-180, involving a path integral for
the gravitational field.} The order of magnitude of the potential
generated by a mass $M$ within the spatial part of the considered
region is $g=\frac{MG}{Lc^2}$, which implies $\Delta
g=\frac{\Delta M G}{Lc^2}$. A comparison with the previous general
expression gives a mass uncertainty $\Delta M =\frac{c^2
L_P}{G}\approx 10^{-5}$ grams, if the time of observations is less
than ${L}/{c}$. On the other side, by allowing an infinite time,
$M$ would be obtained with infinite accuracy. This option,
according to Feynman, could not take place for a mass $M$ fed into
the two-slit apparatus, so that the apparatus will not be able to
uncover the difficulty unless $M$ is at least of order $10^{-5}$
grams. Thus he concluded that
\begin{quote}
Either gravity must be quantized because a logical difficulty
would arise if one did the experiment with a mass of order
$10^{-4}$ grams, or else [...] quantum mechanics fails with masses
as big as $10^{-5}$ grams (Ref. \citenum{ChapelHill}, p. 245).
\end{quote}
Subsequent discussions moved on more formal matters such as the
meaning of the equivalence principle in quantum gravity, with
Feynman contrasting Helmut Salecker's suggestion of a possible
violation of the equivalence principle in the quantum realm. Then
a further remark by Salecker himself brought the participants'
attention back to the topical issue, the necessity of
gravitational quantization. In particular, as explained by an
Editor's note (see Ref. \citenum{ChapelHill}, p. 249), Salecker
hinted to the possibility to build up an action-at-a-distance
theory of gravitation in whole analogy to the electromagnetic
case, with charged quantized particles acting as a source of a
unquantized Coulomb field. At this stage Frederik J. Belinfante
proposed the quantization of both the static part and the
transverse part of the gravitational field (the last one
describing gravitational radiation) as a way to circumvent
difficulties related to the choice of the expectation value of the
stress energy tensor as the source of the gravitational field. As
noticed by Zeh \cite{Zeh}, Belinfante's proposal reflects his
ideas on the ontology of quantum mechanics, which point toward an
epistemic rather than ontic interpretation of the wave function:
\begin{quote}
``There are two quantities which are involved in the description
of any quantized physical system. One of them gives information
about the general dynamical behavior of the system, and is
represented by a certain operator (or operators). The other gives
information about our knowledge of the system; it is the state
vector [...] the state vector can undergo a sudden change if one
makes an experiment on the system. The laws of nature therefore
unfold continuously only as long as the observer does not bring
extra knowledge of his own into the picture.'' (Ref.
\citenum{ChapelHill}, p. 250).
\end{quote}
Belinfante's description used the Heisenberg picture, at variance
with Feynman who reasoned in terms of wave functions as dynamical
objects. Indeed, according to Belinfante ``the wave function [...]
must change for reasons beyond the system's physical dynamics. He
does not refer to ensembles of wave functions or a density matrix
in order to represent incomplete knowledge'' (Ref. \citenum{Zeh},
p. 65).

Feynman promptly replied with his second gedanken experiment, a
Stern-Gerlach experiment with a gravitational apparatus. More in
detail, he considered a spin-$1/2$ particle going through a
Stern-Gerlach apparatus and then crossing the first or the second
of two counters (denoted as $1$ and $2$, respectively), each one
connected by means of a rod to an indicator. He took the indicator
as a little ball with a diameter of $1$ cm, going up or down
depending on the position of the object, at counter $1$ or $2$,
respectively. Quantum mechanics, as underlined by Feynman,
provides in principle an amplitude for the ball up and an
amplitude for the ball down\footnote{Zeh \cite{Zeh} points out how
Feynman's description of the measurement process is very close to
the standard measurement and registration device proposal by John
von Neumann \cite{VonN}.}. However, the ball is chosen to be
macroscopic, so as to be able to produce a detectable
gravitational field, which in turn can move a probe ball. In other
words a channel between the object and the observer is established
via the gravitational field. This ideal experiment led Feynman to
infer the following conclusion about gravity quantization:
\begin{quote}
Therefore, there must be an amplitude for the gravitational field,
provided that the amplification necessary to reach a mass which
can produce a gravitational field big enough to serve as a link in
the chain does not destroy the possibility of keeping quantum
mechanics all the way. There is a bare possibility (which I
shouldn't mention!) that quantum mechanics fails and becomes
classical again when the amplification gets far enough, because of
some minimum amplification which you can get across such a chain.
But aside from that possibility, if you believe in quantum
mechanics up to any level then you have to believe in
gravitational quantization in order to describe this experiment
(Ref. \citenum{ChapelHill}, p. 251).
\end{quote}
A subsequent answer to a question by Hermann Bondi\footnote{``What
is the difference between this and people playing dice, so that
the ball goes one way or the other according to whether they throw
a six or not?''(Ref. \citenum{ChapelHill}, p. 252).} further highlighted,
according to Zeh (Ref. \citenum{Zeh} p. 67), Feynman's position
against the epistemic interpretation of the wave function:
\begin{quote}
I don't really have to measure whether the particle is here or
there. I can do something else: I can put an inverse Stern-Gerlach
experiment on and bring the beams back together again. And if I do
it with great precision, then I arrive at a situation which is not
derivable simply from the information that there is a 50 percent
probability of being here and a 50 percent probability of being
there. In other words, the situation at this stage is not 50-50
that the die is up or down, but \emph{there is an amplitude} that
it is up and an amplitude that it is down -- a complex amplitude
-- and as long as it is still possible to put those amplitudes
together for interference you have to keep quantum mechanics in
the picture. (Ref. \citenum{ChapelHill}, p. 252, our emphasis)
\end{quote}

\section{Wave function collapse}

The aim of this Section is to highlight Feynman's ideas about the
quantum measurement problem. Indeed, from the last sentence of the
last quote, one can infer a clear reference to the problem of wave
function collapse, as well as a hint to decoherence as a possible
solution. A further suggestion can be found in a subsequent
remark:
\begin{quote}
Well, it's a question of what goes on at the level where the ball
flips one way or the other. In the amplifying apparatus there's
already an uncertainty - loss of electrons in the amplifier,
noise, etc. - so that by this stage the information is completely
determined. Then it's a die argument. You might argue this way:
Somewhere in your apparatus this idea of amplitude has been lost.
You don't need it any more, so you drop it. The wave packet would
be reduced (or something). Even though you don't know where it's
reduced, it's reduced. And then you can't do an experiment which
distinguishes  \emph{interfering} alternatives from just plain
odds (like with dice). (Ref. \citenum{ChapelHill}, p. 252, our
emphasis)
\end{quote}
According to Feynman, wave packet reduction occurs somewhere in
his experimental apparatus thanks to the amplifying mechanism, so
that a huge amount of amplification (via the macroscopic
gravitational field of the ball) may be effective in changing
amplitudes to probabilities. Then he wondered about the
possibility to devise an experiment able in principle to avoid the
wave packet reduction due to the amplification process. Subsequent
criticism by Leon Rosenfeld and Bondi further stimulated Feynman's
intuition. So he was led to envisage a sort of quantum
interference in his experiment, driven by the gravitational
interaction between macroscopic balls, which could be described by
means of a quantum field with suitable amplitudes taking a value
or another value, or to propagate here and there. But, as Bondi
suggested, any irreversible element should be removed, such as for
instance the possibility of gravitational links to radiate.
Probably, this is another point in which a further hint to the
role of decoherence in destroying quantum interference can be
recognized, even if in Bondi's and Feynman's words a reference is
made only to classical irreversibility \cite{Zeh}. In fact the
meaning of decoherence, its origin and its role in smearing out
phase relations as well as in triggering the transition to
classicality, were still unclear in the late 1950s\footnote{For a
historical and research account on decoherence, see Ref.
\citenum{ZurekRev} and references therein}. The discussion went on
with Feynman arguing that quantum interference might eventually
take place with a mass of macroscopic size, say about $10^{-5}$
gram or $1$ gram, and hinting to the possible role of gravity in
destroying quantum superpositions. In his words:
\begin{quote}
There would be a \textit{new principle}! It would be
\textit{fundamental}! The principle would be: -- \textit{roughly}:
\textit{Any piece of equipment able to amplify by such and such a
factor} ($10^{-5}$ grams or whatever it is) \textit{necessarily
must be of such a nature that it is irreversible}. It might be
true! But at least it would be fundamental because it would be a
new principle. There are two possibilities. Either this principle
-- this missing principle -- is right, \textit{or} you can amplify
to any level and still maintain interference, in which case it's
absolutely imperative that the gravitational field be quantized...
\textit{I believe}! \textit{or} there's another possibility which
I have't thought of (Ref. \citenum{ChapelHill}, pp. 254-255,
emphasis in original).
\end{quote}
The same ideas would have been pursued later by Feynman in his
Lectures on Gravitation\cite{Feynman:1996kb}, where he considered
``philosophical problems in quantizing macroscopic objects'' and
hinted to the possibility of a failure of quantum mechanics
induced by gravity:
\begin{quote}
I would like to suggest that it is possible that quantum mechanics
fails at large distances and for large objects. Now, mind you, I
do not say that I think that quantum mechanics \textit{does} fail
at large distances, I only say that it is not inconsistent with
what we do know. If this failure of quantum mechanics is connected
with gravity, we might speculatively expect this to happen for
masses such that $\frac{GM^2}{\hbar c}=1$, of $M$ near $10^{-5}$
grams, which corresponds to some $10^{18}$ particles (Ref.
\citenum{Feynman:1996kb}, pp. 12-13).
\end{quote}
Within the same set of lectures (Ref. \citenum{Feynman:1996kb}, p.
14), the possibility is recognized that amplitudes may reduce to
probabilities for a sufficiently complex object, thanks to a
smearing effect on the evolution of the phases of all parts of the
object. Such a smearing effect could have a gravitational origin.
A similar idea is expressed in the end of the letter to Weisskopf
\cite{WeisskopfLetter}, as already mentioned in the Introduction.
This shows how Feynman was open-minded with respect to all
possibilities, despite his strong belief in the quantum nature of
reality.

The debate on gravity quantization here highlighted and, in
particular, Feynman's deep insights, triggered subsequent research
on the possibility of a gravity-induced collapse of wave function,
as a viable solution to the measurement problem in quantum
mechanics
\cite{karol,frenkel,frenkel1,diosi0,diosi,diosi1,penrose0,penrose1,penrose2,penrose3}.
The idea is attractive since gravity is ubiquitous in nature, and
gravitational effects depend on the size of objects, so it has
been greatly developed in the subsequent years up to the present
day. Along this line of thinking, Roger Penrose suggested that a
conflict emerges when a balanced superposition of two separate
wave packets representing two different position of a massive
object is considered \cite{penrose0,penrose1}. In his words (see
Ref. \citenum{penrose4}, p. 475): ``My own point of view is that
as soon as a \textit{significant} amount of space-time curvature
is introduced, the rules of quantum linear superposition must
fail''. Clearly the conflict is the result of putting together the
general covariance of general relativity and the quantum
mechanical superposition principle. By assuming in each space-time
the validity of the notions of stationarity and energy,  and by
taking the difference between the identified time-translation
operators as a measure of the ill-definiteness of the final
superposition's energy, the decay time for the balanced
superposition of two mass distributions can be estimated:
\begin{eqnarray}\label{PenroseEnergy}
t_D=\frac{\hbar}{\Delta E_{grav}}.
\end{eqnarray}
Here $\Delta E_{grav}$ is the gravitational self-energy of the
difference between the mass distributions of each of the two
locations of the object. This means that massive superpositions
would immediately undergo collapse. The same idea led Penrose to
introduce the so called Schr\"{o}dinger-Newton equation
\cite{penrose2}, which governs a peculiar non-linear evolution of
the center-of-mass wave function. Its soliton-like dynamics is the
result of a competition between a partial shrinking due to the
non-linear term and a partial spreading due to the usual dynamical
term.

More recently, other collapse models(called \textit{dynamical}
collapse models) have been put forward
\cite{random1,random2,random3,random4}, where the collapse of the
wave function is induced by a different mechanism, i.e. the
interaction with a random source such as an external noise source.

From the experimental side, a lot of proposals have been made as
well, mainly aimed to explore a parameter range where both quantum
mechanics and gravity are significant. For instance it has been
possible to create a quantum superposition state with complex
organic molecules with masses of the order $m=10^{-22}$ kg
\cite{arndt1,arndt2}. Recent experimental proposals involve
matter-wave interferometers \cite{matterW1,matterW2,matterW3},
quantum optomechanics \cite{bose,marshall,mio,oriol} and
magnetomechanics \cite{mag1}. Today a major challenge is, on one
hand, to design viable experiments at the interface between the
quantum and classical worlds, while on the other hand to reveal
and discriminate gravity-induced decoherence from environmental
decoherence in such experiments. However, despite such a huge
theoretical and experimental effort, a definite answer to the
quantum measurement problem as well as to the problem of the
emergence of classical from the quantum world is still lacking.

\section{The role of the observer in a closed Universe}

In this Section we deal with the issue of the role of the observer
in a closed Universe, as emerged from discussions carried out in
the closing session of Chapel Hill conference. Here Wheeler gave
the first public presentation of Everett's relative-state
interpretation of quantum mechanics \cite{Everett,Osnaghi}, as
follows:
\begin{quote}
General relativity, however, includes the space as an integral
part of the physics and it is impossible to get outside of space
to observe the physics. Another important thought is that the
concept of eigenstates of the total energy is meaningless for a
closed Universe. However, there exists the proposal that there is
one ``universal wave function". This function has already been
discussed by Everett, and it might be easier to look for this
``universal wave function" than to look for all the propagators
(Ref. \citenum{ChapelHill}, p. 270).
\end{quote}
Feynman promptly replied by characterizing of Everett's approach
as ``many-worlds":
\begin{quote}
The concept of a ``universal wave function" has serious conceptual
difficulties. This is so since this function must contain
amplitudes for all possible worlds depending on all
quantum-mechanical possibilities in the past and thus one is
forced to believe in the equal reality of an infinity of possible
worlds (Ref. \citenum{ChapelHill}, p. 270).
\end{quote}
The same idea will be presented some years later in the Lectures
on Gravitation (Ref. \citenum{Feynman:1996kb}, pp. 13-14), where
Feynman also discussed the role of the observer in quantum
mechanics. The Schr\"{o}dinger's cat paradox allowed him to
illustrate the difference between the results of a measurement
carried out by an external as well as an internal observer. While
the external observer describes his results by an amplitude, with
the system collapsing into a well-defined final state after the
measurement, according to the internal observer the results of the
same measurement are given by a probability. Clearly the absence
of an external observer leads to a paradox, which is much more
effective when considering the whole Universe as described by a
complete wave function. This Universe wave function is governed by
a Schr\"{o}dinger equation and implies the presence of an infinite
number of amplitudes, which bifurcate from each atomic event. In
other words, according to Feynman, the Universe is constantly
spitting into an infinite number of branches, as a result of the
interactions between its components. Here the key observation is
that interactions play the role of measurements. As a consequence,
an inside observer knows which branch the world has taken, so that
he can follow the track of his past. Feynman's conclusion brings
into play a conceptual problem:
\begin{quote}
Now, the philosophical question before us is, when we make an
observation of our track in the past, does the result of our
observation become real in the same sense that the final state
would be defined if an outside observer were to make the
observation (Ref. \citenum{Feynman:1996kb}, p. 14)?
\end{quote}
A further discussion of the meaning of the wave function of the
Universe is carried out later in the Lectures on Gravitation (Ref.
\citenum{Feynman:1996kb}, pp. 21-22). Here Feynman restated his
``many-worlds" characterization of Everett's approach with
reference to a ``cat paradox on a large scale'', from which our
world eventually could be obtained by a ``reduction of the wave
packet". Concerning this reduction, he wondered about the relation
between Everett's approach and collapse mechanisms of whatever
origin. Perhaps it may be relevant to quote a comment by John P.
Preskill, within a talk about the Feynman legacy, given at the APS
April Meeting (cf. Ref. \citenum{PreskillTalk}, slide 29):
\begin{quote}
When pressed, Feynman would support the Everett viewpoint, that
all phenomena (including measurement) are encompassed by unitary
evolution alone. According to Gell-Mann, both he and Feynman
already held this view by the early 1960s, without being aware of
Everett's work\footnote{We do not agree that Feynman was not aware
of Everett's work in the 1960s, since he commented on it at Chapel
Hill in 1957.}. However, in 1981 Feynman says of the many-worlds
picture: ``It's possible, but I am not very happy with it".
\end{quote}

Historically, Everett's proposal \cite{Everett} was the first
attempt to go beyond the Copenhagen interpretation in order to
apply quantum mechanics to the Universe as a whole. To this end,
the well known separation of the world into ``observer" and
``observed" has to be superseded by promoting the observer to be
part of the system, while the usual quantum rules are still
effective in measuring, recording or doing whatever operation.
Quantum fluctuations of space-time in the very early Universe have
also to be properly taken into account. On the other hand, this
proposal lacks an adequate description of the origin of the
quasi-classical realm as well as a clear explanation of the
meaning of the branching of the wave function.

Everett's work has been further developed by many authors
\cite{dh1,dh2,dh3,dh4,dh5,dh6} and its more recent generalization
is known as \textit{decoherent histories approach to quantum
mechanics of closed systems}. A characterizing feature of this
formulation is that neither observers nor their measurements play
a prominent role. Furthermore the so called \textit{retrodiction},
namely the ability to construct a history of the evolution of the
Universe towards its actual state by using today's data and an
initial quantum state, is allowed. The process of prediction
requires to select out decoherent sets of histories of the
Universe as a closed system, while decoherence in this context
plays the same role of a measurement within the Copenhagen
interpretation. Decoherence is a much more observer-independent
concept and gives a clear meaning to Everett's branches, the main
issue being to identify mechanisms responsible for it. In
particular, it has been shown that decoherence is frequent in the
Universe for histories of variables such as the center of mass
positions of massive bodies \cite{Joos}.

\section{Concluding remarks}

In this paper we have retraced Feynman's thoughts on foundational
issues in quantum mechanics. Such ideas were rarely expressed by
him, who mainly linked his name to his masterful application of
thus theory, but they clearly emerged whenever he focused on the
problem the interface of quantum mechanics and gravity, with the
ensuing great conceptual questions. These problems, such as that
of the collapse of the wave function, of macroscopic
superpositions and of observers which are parts of the observed
system itself, are of course not limited to the gravitational
realm, but lie at the very heart of quantum physics. Despite much
progress having been achieved since Feynman's times, part of which
had been anticipated by him and other participants to the Chapel
Hill conference, much work still needs to be done, before his most
famous remark\cite{Messenger} ``I think I can say that nobody
understands Quantum Mechanics'' can be considered to be no longer
true.




\end{document}